\documentclass[aps,prb,preprint,showpacs]{revtex4}
\usepackage{preamblepaper}
\setcitestyle{numbers,square}
\begin{document}
\title{A phenomenological theory of the optical magnetization reversal}

\author{Marco Menarini}
\email{menarini.marco@gmail.com}
\author{Vitaliy Lomakin}%
\affiliation{%
 Department of Electrical and Computer Engineering, Center for Memory and Recording Research, University of California, San Diego, La Jolla, California 92093-0319}%

\date{\today}

\begin{abstract}

All-optical switching of the magnetization in magnetic nanostructures by femtosecond circularly polarized laser pulses has been demonstrated in several systems. We present a Landau-Lifshitz-Lambda (LLL) model which describes the magnetization dynamics using three density states: two ferromagnetic grounds states and an excited optical state. One of the ferromagnetic ground states is optically excited by circularly polarized light to a spin reversed state, which is then “Coulomb collapsed” to the magnetization reversed ground state. The time evolution of the optically excited states is described by a Lindblad master equation, in which the optical excitation is introduced via the Hamiltonian. Dissipation terms are introduced via Lindblad operators. The LLL model combines the precessional motion of the magnetization described by the Landau-Lifshitz theory, with the response of the three level $\Lambda$ system. The optical excitation lasts for the duration of the laser pulse and the system relaxes at a fast rate due to the electron-electron interaction.
We study the solution of the eigenvalues problem of the optical equation of motion for the magnetization and identify a coherent and incoherent regimes and derive an LLL model that can be integrated with existing micromagnetic codes to describe optical excitation of magnetic materials.

\end{abstract}
\maketitle


\section{\label{sec:introduction}Introduction}

Ultrafast magnetization reversal induced by optical excitation without an external magnetic field, known as all-optical switching (AOS), has been a topic of intense research in recent years for its importance in understanding the magnetization dynamics beyond the usual magnetization dynamics time scale and in potential applications to memory storage. There is a variety of interpretations of the causes of the underlying physical processes. Two comprehensive reviews \cite{Bigot2013, Kirilyuk2010} both listed two groups of the causes, (i) the laser induced heating and (ii) the direct optical driving of the magnetization, including the optical pumping of the electronic transitions under the spin-orbit interaction and the inverse Faraday Effect (IFE). A third cause, the “toggle” mechanism, emphasized by another review \cite{Kirilyuk2013}, applies to AOS of ferrimagnets, occurring in the antiferromagnetically aligned bipartite rare-earth and transition metal sublattices.

The first group of mechanisms, labeled laser heating mechanism for the magnetization reversal, has been used to explain many experiments and theories, such as Ref. \cite{Ostler2012}. A confirmation of the role of heat is the demonstration of the single-shot reversal ascribed to the heat flow \cite{Gorchon2017}. The non-equilibrium thermal approach has also been extended to the magnetic circular dichroism (MCD) \cite{Ellis2016, Khorsand2012}, the difference in absorption between the left and right circularly polarized light, to explain AOS in ferromagnets. However, the strength of the MCD effect required in simulations appears to be greater than what has been observed experimentally \cite{Medapalli2017}.

The second group, the coherent optical driving of the magnetization, was shown by optical pump and probe Kerr and Faraday experiments \cite{Bigot2009}. The light polarization dependence of the magnetization reversal led to the interpretation of polarization photon driving the electron spin reversal \cite{Vahaplar2012} and is found in a wide range of other materials \cite{ElHadri2016, Lambert2014, Medapalli2017, Takahashi2016}. IFE, a phenomenon demonstrated by experiment \cite{Holzrichter1971, VanderZiel1965}, is a special case of this angular momentum transfer mechanism from photon to electron spin, confined to the second order of the oscillating electric field \cite{Shen1984}. This may explain why the effective magnetic field that IFE is expected to produce and the duration that it could apply through the ultrafast laser pulse are both inadequate to explain the AOS experiments \cite{Ellis2016}. 
An example of a coherent driving mechanism is given in the theory of the optical driving of a sublattice magnetization in the antiferromagnetic NiO  \cite{Lefkidis2009} by combining the many-electron spin interaction physics with coherent optics. NiO, being an insulator with an energy gap, favors coherent optical processes. On the other hand, for a ferromagnetic metal, where a magnetization state is in an energy continuum with magnon states and with charged excited states, the magnetization is driven primarily by an incoherent optical excitation \cite{Bigot2009}. Laser heating and optical pumping are both in the incoherent driving class. The physical nature of the excitation and transition between a coherent and incoherent driving mechanisms is described later in building our theory.

In the third group, the toggle mechanism for ferrimagnets, the laser pulse first drives the sublattice with lower magnetocrystalline anisotropy energy to be reversed to parallel to the one with the higher anisotropy energy and then the instability of the spin structure causes the second sublattice to the reversed antiparallel alignment\cite{Atxitia2013, Radu2011}. It is perhaps reasonable to include as the toggle action, the driving of a ferromagnet layer coupled to an antiferromagnet layer by the exchange bias \cite{Vallobra2017}. 

The laser heating mechanism for ultrafast magnetization dynamics is in the theories of the first group in a variety of magnetic pathways, such as micromagnetic modeling of ferrimagnetic materials \cite{Atxitia2012} or ferromagnetic materials subject to sub-picosecond thermal pulses \cite{Mendil2014},  and the toggle process in ferrimagnets \cite{Ostler2012}. A quantum theory of IFE has been given \cite{Battiato2014, Zhang2000} that will facilitate the theory of the optical driving of magnetization reversal based on the second order processes. Theories of demagnetization include (i) the requirement of both laser excitation and spin-orbit interaction\cite{Cywinski2007, Zhang2000}; (ii) the demonstration that the exchange interaction between the localized electrons which provide the magnetization and the carrier electrons which can be optically excited as the cause \cite{Cywinski2007}; (iii) a two-level system model study of the optical driving of the spin polarization and the spin-orbit effects \cite{Zhang2008}. The theories for demagnetization may be extended to apply to optical reversal with the appropriate treatment of the optical excitation, as shown by Gridnev \cite{Gridnev2013}. The theory of the ferrimagnet with localized electrons and itinerant ones forming two ferromagnetic states antiferromagnetically aligned has given an explanation of the toggling effect \cite{Baral2015, Gridnev2016}. The thermal and non-thermal processes in AOS was posed as an either-or proposition in the case of \ch{GdFeCo} \cite{Zhang2013}, but in general the two consequences of electron state excitation and of heating by the laser could both contribute to magnetization reversal, as shown by Ref. \cite{Gridnev2013}. The theories briefly reviewed above form the basis of our model construction.

The fundamental question of the optical driving of the magnetization reversal is the speed much higher than the natural precession speed of the macro-spin state and spin wave dynamics in the presence of the laser heating. We present a theory and simulation to understand the effects of the simplest underlying optically driving and the fast relaxation to the magnetic ground states. We aim for this theory of excitation and relaxation sequence to apply to the wide range of materials subjected to optical magnetization with polarized light \cite{Lambert2014, Takahashi2016}. 

We propose a three-state model, referred to as the Landau-Lifshitz-Lambda (LLL) model, of a ferromagnetic ground state driven by the circularly polarized laser light to an excited state of reversed magnetization, which relaxes by fast electronic means to the ground state of the reversed magnetization \cite{Beaurepaire1996}. The   shape of the three states is similar to the states in the Raman process but differs in the pump-and-relax process compared to the two-beam optical process. Each macrospin state is associated with proximate bands with energy of the charge and spin wave states including the effects of spin-orbit interaction and crystalline anisotropy effects. We approximate the bands near the three states by probability distributions to account for the fluctuation and dissipation effects as the thermal and stochastic effects on the magnetization and optical polarization dynamics \cite{Evans2012, Garanin1997}. 

\section{\label{sec:model}Lambda model for optical pumping in magnetic systems}

Our aim is to build a model of the physical process of magnetization reversal, which (i) is driven by optical excitation and electronic demagnetization, (ii) to which the thermal effects(i.e. laser heating) may be added, and (iii) ready for numerical simulation. In this section, we concentrate on the first step. The model contains three key states, two ferromagnetic ground states separated by uniaxial anisotropy and an optically excited state. 
The circularly polarized laser pulse excites the initial ferromagnetic ground state to an excited state with energy far above the anisotropy potential barrier between the two ground states. A sufficient amount of spin reversal, due to spin-orbit coupling, biases the state against reverting back to the original magnetic ground state. The excited state decays by the spin-conserved electron-electron interaction fast electron processes (i.e. Coulomb collapse) to the low-energy excited states in the anisotropy energy valley of the reversed magnetization ground state. The decay to the final ground state is dominated by interaction with phonons and magnons. The optical excitation leg is reasoned to be supported by the extant experiments and theories and the fast decay leg supported by the electronic causes of the demagnetization process, both of which are incorporated in this section. The model is tested to a certain extent by the numerical simulation in the following sections.

In a ferromagnet, the proximate energy above the two ground states and around the optically excited state are the charge and spin states, the latter including the magnons in the Heisenberg model or the spin-flip excitation in the Stoner model \cite{Galanakis2012,Glazer1984}. This scenario is equivalent to rendering the skeleton three states to an open system. Thus, the system may be treated by the quantum dynamics of a standard three-level system \cite{Scully1997}, governed by coherent and incoherent optical excitation with the upper level subject to a fast dissipation to the reversal state. The continua around these three states are represented by three probability distributions of the three states to provide the fluctuation effects of the magnetic sector and of the optical sector. The equations of motion with fluctuations are equivalent to the modified Bloch equations for the two-level system \cite{Berman1985,Yamanoi1984}.

\subsection{\label{sec:basis} The basis macrospin states}

\begin{figure}
    \centering
    \includegraphics{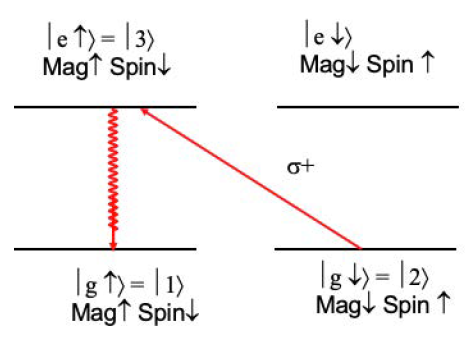}
    \caption{The horizontal lines represent the macrospin states. The labels $\stgd$ and  $\stgu$ denotes the magnetization-down and magnetization-up ground state, respectively. The label $\steu$  denotes  the magnetization-up excited state. The optical transition from the ground state to the excited state is driven by the $\lhc$ light. The relaxation from   $\steu$   to  $\stgu$  is a fast-non-spin flip process, which is a non-radiative Coulomb interaction induced decay. The fourth state $\sted$ is present to indicate a possible pathway for the  light.}
    \label{fig:scheletonprocess}
\end{figure}

\begin{figure}
    \centering
    \includegraphics{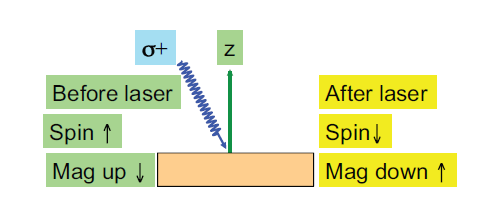}
    \caption{Following the traditional notation, the z axis is defined to be the growth axis of the film. Vectors along the positive z axis direction is said to be “up”, such as the light wave vector, magnetization, or spin (total angular momentum in the case of spin orbit interaction). Vectors along the negative z axis are said to be “down”. The $\lhc$  circular polarization of light propagating downwards, therefore, flips the electron spin from up to down. The magnetization direction is opposite to that of the electron spin because of its negative electronic charge $-e$  in the magneton factor $-e\hbar/2m$ .}
    \label{fig:mechanism}
\end{figure}

The three key basis states of all the spins of the system, shown in \cref{fig:scheletonprocess}, are   $\sta$,   $\stb$, the two macrospin ferromagnetic ground states in a uniaxial anisotropy system and  $\stc$  an optically excited state. State $\stb$ is chosen to be the initial magnetization state which can be excited by a right-handed ($\lhc$ ) circularly polarized light to an excited state $\stc$ of reversed magnetization, which decays without change in magnetization direction into the reversal state $\sta$. For the convenience of connecting the theory to possible experiments, the actual spin and magnetization labels of vertical arrows are specified in \cref{fig:mechanism} for a particular optical process. It shows the relation of the helicity of light propagating against the growth direction of the ferromagnetic film, along and against which the respective directions of the magnetization and spin are commonly defined.

Because the $\Lambda$  system is open, subject to control and dissipation, we represent its mixed state by a density operator $\dens$. We find it useful to transform the matrix elements to the defined Bloch vectors for pairs of states, in particular, the magnetization vector $\vecm$ between the magnetization states $\sta$,$\stb$, the optical polarization $\vecp$ between the optically connected states  $\stb$,$\stc$ ,  and the charge polarization without spin change $\vecd$ between the states $\sta$,$\stc$ . Thus, the vectors in terms of the density matrix $\dens$ are given by,
\begin{equation}
    \label{eq:densitymatrix}
    \dens=\begin{bmatrix}
        \rho_{11} & \rho_{12} & \rho_{13} \\
        \rho_{21} & \rho_{22} & \rho_{23} \\
        \rho_{31} & \rho_{32} & \rho_{33} 
    \end{bmatrix}=\frac{1}{2}
    \begin{bmatrix}
        1-\vdz & \vmm & \vdp \\
        \vmp & 1-\vpz & \vpp \\
        \vdm & \vpm & \vpz+\vdz 
    \end{bmatrix} \, .
\end{equation}
We follow the convention of the spin and optics communities and call a diagonal terms the \textquote{\emph{populations}}  of a basis state and the off-diagonal terms the  \textquote{\emph{coherences}} between two basis states. The transverse components of the magnetization are contained in
\begin{equation}
    \label{eq:coherence}
    \vmpm = \vmx\pm i\vmy \, ,
\end{equation}

and similarly for $\vppm$ and $\vdpm$. The unit trace of the density matrix constrains the choice of the longitudinal components to two,
\begin{align}
    \label{eq:populations}
    \vdz \coloneqq \rho_{33}-\rho_{11}+\rho_{22} \, , \\
    \vpz \coloneqq \rho_{33}-\rho{22}+\rho_{11} \, ,
\end{align}
which we refer to, respectively, as charge polarization and optical polarization.

To get a physical sense of the new polarization vector components $\left(\vpz,\vdz\right)$, we consider the behaviour of the density submatrix of each pair of states, named a sector. The optical sector of states $\stb$ and $\stc$, and the charge sector of states $\sta$ and $\stc$, as the two legs of the $\Lambda$ system, have a limited amount of symmetry. Thus, the two polarization vectors $\vecp$,$\vecd$ can be switched, but the optical polarization population $\vpz$ has spin change, while the charge polarization  population $\vdz$ does not. When $\vdz=1$, state $\sta$ is empty and the states $\stb$ and $\stc$ of the optical sector become an isolated two-state system, with the normal appearance of the two-level system,
\begin{equation}
    \label{eq:twolevels}
    \begin{bmatrix}
    \rho_{22} & \rho_{23} \\
    \rho_{32} & \rho_{33} 
    \end{bmatrix}
    =
    \begin{bmatrix}
    1-\vpz & \vpp \\
    \vpm & \vpz+1 
    \, .
    \end{bmatrix}
\end{equation}
Similarly, when $\vpz=1$, the charge polarization vector $\vecd$ between states $\sta$ and $\stc$ serves the same role but without the spin flip. We will see later that the Hamiltonian will make the charge sector serve as a leakage conduit. In the magnetic sector of states $\sta$ and $\stb$, the natural definition of the $z$-component of the magnetization vector is, from \cref{eq:densitymatrix},
\begin{equation}
    \label{eq:magnetizationpopulation}
    \vmz  \coloneqq \rho_{11}-\rho_{22}=\frac{1}{2}\left(\vpz-\vdz\right)
    \, .
\end{equation}

The density operator is a Hermitian operator subject to two restrictions: (i) it is positive (easily tested by its eigenvalues be $\lambda_i\ge0$ ); (ii) its trace is $\Tr\left({\dens}\right)=1$ . It follows that $\Tr\left({\dens^2}\right)\le1$ , where the equality holds if and only if the state is pure (i.e. $\rho_{ij}=0$ for $i\neq j$). In a two-state system, the Bloch vector is confined to a sphere with its surface defined by $\Tr\left({\dens^2}\right)=1$  for pure states. However, in systems with more than two states, the corresponding hypersphere does not present a dividing hypersurface for valid representation of states. The issue is discussed in Appendix \ref{sec:app_vector}. The second issue is that the mapping between the density matrix elements and the polarization (optical and magnetic) is not a one-to-one mapping \cite{Bertlmann2008}. An obvious example is the two dimensional submatrices, where spin $1/2$  rotation has a period of  $4\pi$  and not $2\pi$  \cite{Werner1975}, arising from the phase terms in the non-diagonal elements. Because the three key states are macrospin states, we may consider the mapping one-to-one, neglecting the subtle quantum features such as the spin $1/2$ case, whence the complete rotation of the macro-spin polarization vector is $2\pi$.

\subsection{\label{sec:motion} The equation of motion}

We model the dynamics of the magnetization as that of the system of three most probable or average states. The continua surround the respective macrospin states and they are treated as an environment that makes the dynamics of an open system. Therefore, the macrospin equations of motion contains coherent driving terms and dissipative terms. A formal method to construct such semigroup evolution operators is by expressing the master equation of the reduced density matrix for the macrospin state in terms of both the unitary evolution generated by the Hermitian Hamiltonian $\hham$ (conservative terms) and the Lindblad operators $\lindblad{k}$ (dissipative terms) in the Lindblad equation \cite{Lindblad1976},
\begin{equation}
    \label{eq:lindblad}
    \deriv{\dens}{t}=-i\commut{\hham}{\dens}+\sum_{k}{\left(\lindblad{k}\dens\lindblad{k}^\dagger-\dfrac{1}{2}\acommut{\lindblad{k}^\dagger\lindblad{k}}{\dens}\right)}
    \, .
\end{equation}
The first term on the right hand side is the unitary dynamics in the rotating wave approximation driven by the Hamiltonian in units of frequency and in the basis set of $\sta$, $\stb$, $\stc$,
\begin{equation}
    \label{eq:hamiltonian}
    \hham=\begin{bmatrix}
    \vbz & \vbm & 0 \\
    \vbp & -\vbz & \frac{i\Omega}{2} \\
    0 & -\frac{i\Omega}{2} & \Delta-\vbz 
    \end{bmatrix}
    \, .
\end{equation}
Here, $\vecb=\gamma\vech$, where $\gamma$ is the gyromagnetic ratio and $\vech$ is the effective magnetic field acting on the magnetization, than includes the contribution from the applied, anisotropic, and mean-field approximation of the micromagnetic exchange field \cite{Garanin1997}. The laser detuning $\Delta=\omega_L-\omega_0$ is given by the difference between the effective laser frequency $\omega_L$ and the resonance frequency $\omega_0$ between states $\stb$,$\stc$.  The Rabi frequency $\Omega$ for the $\lhc$ light is defined as:
\begin{equation}
    \label{eq:rabifreq}
    \Omega=2\mue\ef\, ,
\end{equation}
where $\mue$ is the transition dipole moment and $\ef$ is the electric field. All energy terms are in units of inverse fs. The spin flip in the optical channel is due to the transfer of angular momentum between the spin-orbit mixed states by the circularly polarized light.

The second term on the right end side of \cref{eq:lindblad} is composed of the dissipative energy terms given in terms of the Lindblad operators $\lindblad{k}$'s. The terms are the phenomenological representations of the effects of the
continua of states around the three chosen macrospin states in our model. Details of the construction of these operators and equations are given in Appendix \ref{sec:app_motion}.

When the density matrix is converted to the eight components of the polarization and magnetization, the equations are,
\begin{flalign}
    \label{eq:lll_pz}
    &\deriv{\vpz}{t}= +
    \Omega\vpx+2\left(\vbx\vmy-\vby\vmx\right)
    \, , \\
    \label{eq:lll_dz}
    &\deriv{\vdz}{t}=-
    2\left(\vbx\vmy-\vby\vmx\right)
    -\ccol\left(\vpz+\vdz\right)
     \, , \\
    \label{eq:lll_px}
    &\deriv{\vpx}{t}=-
    \Delta\vpy-\dfrac{1}{2}\Omega\left(2\vpz+\vdz-1\right)+\left(\vbx\vdy+\vby\vdx\right)
    -\dfrac{1}{2}\ccol\vpx
    -\pdeph\vpx  \, , \\
    \label{eq:lll_py}
    &\deriv{\vpy}{t}=+
    \Delta\vpx-\left(\vbx\vdx-\vby\vdy\right)
    -\dfrac{1}{2}\ccol\vpy
    -\pdeph\vpy \, , \\ 
 \label{eq:lll_dx}
    &\deriv{\vdx}{t}=+
    \left(2\vbz-\Delta\right)\vdy
    +\dfrac{1}{2}\Omega\vmx
    +\left(\vbx\vpy-\vby\vdx\right)
    -\dfrac{1}{2}\ccol\vdx
    -\pdeph\vdx \, , \\
    \label{eq:lll_dy}
    &\deriv{\vdy}{t}=-
    \left(2\vbz-\Delta\right)\vdx
    -\dfrac{1}{2}\Omega\vmy
    -\left(\vbx\vpx+\vby\vdy\right)
    -\dfrac{1}{2}\ccol\vdy
    -\pdeph\vdy \, , \\ 
 \label{eq:lll_mx}
    &\deriv{\vmx}{t}= -
    2\left(\vby\vdz+\vbz\vmy\right)
    +\vby\left(\vpz+\vdz\right)
    -\dfrac{1}{2}\Omega\vdx \, ,\\
    \label{eq:lll_my}
    &\deriv{\vmy}{t}=+
    2\left(\vbz\vmx+\vbx\vdz\right)
    -\vbx\left(\vpz+\vdz\right)
    +\dfrac{1}{2}\Omega\vdx  \, .
\end{flalign}
Each equation is arranged in the sequence of contributions to the rate on the left side in order of terms with coherent driving factors of $\Omega$, $\Delta$, $\vecb$, the coulomb collapse rate $\ccol$, and the pure dephasing contribution $\pdeph$. To study the magnetization dynamics, it is useful to express the model explicitly in terms of the population of the magnetization sector instead of the charge and optical sectors. To do so, we can recast the rate equation for the magnetization sector given in \cref{eq:lll_mx,eq:lll_my} using the relationship given in \cref{eq:magnetizationpopulation},
\begin{flalign}
    \label{eq:lll_mz}
    &\deriv{\vmz}{t}= +2\left(\vbx\vmy-\vby\vmx\right)
    +\dfrac{\Omega}{2}\vpx
    +\ccol\left(\vpz-\vmz\right)
    \, , \\
 \label{eq:lll_mx1}
    &\deriv{\vmx}{t}= +2\left(\vby\vmz-\vbz\vmy\right)
    -\dfrac{\Omega}{2}\vdx
    \, ,\\
    \label{eq:lll_my1}
    &\deriv{\vmy}{t}=-2\left(\vbx\vmz-\vbz\vmx\right)
    +\dfrac{\Omega}{2}\vdy
    \, .
\end{flalign}
The magnetic relaxation times left out of the equations can be re-inserted by means of the Lindblad operators in the magnetic sector.

In the optical excitation from state $\stb\rightarrow\stc$, the localized orbital of an electron in the macrospin state goes from d (or f )  to p (or d), respectively, with an angular momentum decrease of one $\hbar$ by the  $\lhc$  photon traveling against the spin axis as in \cref{fig:mechanism}, via the optical dipolar interaction. The optical process by itself does not flip the electron spin. But in the presence of spin-orbit coupling, mixed spin-up and spin-down states are present. The $\lhc$ photon transfers a unit of total angular momentum, and the selection rule becomes $\Delta m_j=+1$. This causes the spin-flip of the optically involved states \cite{Cywinski2007, Zhang2008b}. In fact, such selection rule allows for transitions $\stgd\rightarrow\steu$, but not $\stgu\rightarrow\sted$, resulting for an effective spin-flip state that is described by the optical excitation via the Rabi frequency between $\stb$ and $\stc$ (\cref{eq:rabifreq}), which is proportional to the optical electric field $\ef$.
Similar symmetry breaking effects has been shown in experiment \cite{Khorsand2012}, and can be explained in theory \cite{Gridnev2016} by time-reversal symmetry breaking of the ferromagnetic material. 

In the system of \cref{eq:lll_pz,eq:lll_dz,eq:lll_px,eq:lll_py,eq:lll_dx,eq:lll_dy,eq:lll_mx,eq:lll_my}, $\pdeph$ only appears in the coherence of $\vecd$ and $\vecp$, perpendicular to the applied laser direction z. The relaxation rate of the population $\vdz$ and $\vdz$ (and implicitly $\vmz$) only depend on the decay rate $\ccol$.

The electron relaxation process also plays an important role in the sub-picosecond demagnetization under fast optical excitation, as found in a variety of fast optics experiments \cite{Cheskis2005}. The loss of the magneto-optic Kerr effect (MOKE) contrast of the remnant magnetization saturating at high excitation densities was associated with an instantaneous “Stoner gap collapse” in the same paper. This fast relaxation is explained by an electron scattering theory \cite{Krauss2009} and computations using the density functional theory, which includes the electron-electron interaction in the spin polarized configuration \cite{Zhang2015}.

\section{\label{sec:optical} Study of Optical Excitation}

In this section, our aim is to demonstrate the key combination of the optical excitation and the electron driven decay in AOS observed in ferromagnetic material. We focus on the \textquote{\emph{resonant optical model}}, where the light is considered to be of a single transition frequency ($\Delta=0$) and no external field is applied ($\vecb=0$). The optical excitation is modeled as a continuous wave source turned on at $t=0$ such that $\Omega(t)=\Omega_0$ for $t>0$.

Assuming the system initially is in the $\stb$ (i.e. $\rho_{22}=1$), only the population channels and the coherence component of the optical sector $\vpx$ are excited by $\Omega$. We write the system of  \cref{eq:lll_pz,eq:lll_dz,eq:lll_px,eq:lll_py,eq:lll_dx,eq:lll_dy,eq:lll_mx,eq:lll_my} for the resonant optical model as,
\begin{flalign}
    \label{eq:lll_rom_dz}
    &\deriv{\vdz}{t}=-\ccol\left(\vpz+\vdz\right)
     \, , \\
    \label{eq:lll_rom_pz}
    &\deriv{\vpz}{t}= +\Omega\vpx
    \, , \\
    \label{eq:lll_rom_px}
    &\deriv{\vpx}{t}=
    -\dfrac{1}{2}\Omega\left(2\vpz+\vdz-1\right)
    -\dfrac{1}{2}\ccol\vpx
    -\pdeph\vpx  \, ,
\end{flalign}
or, equivalently, in terms of the magnetization and the optical sector, using the relationship \cref{eq:magnetizationpopulation} as, 

\begin{flalign}
     \label{eq:lll_rom_mz1}
    &\deriv{\vmz}{t}= +\dfrac{\Omega}{2}\vpx
    +\ccol\left(\vpz-\vmz\right)
    \, , \\
    \label{eq:lll_rom_pz1}
    &\deriv{\vpz}{t}= +\Omega\vpx
    \, , \\
    \label{eq:lll_rom_px1}
    &\deriv{\vpx}{t}=
    -\dfrac{1}{2}\Omega\left(3\vpz-2\vmz-1\right)
    -\dfrac{1}{2}\ccol\vpx
    -\pdeph\vpx  \, .
\end{flalign}
We can express the system of \cref{eq:lll_rom_pz,eq:lll_rom_dz,eq:lll_rom_px} (or equivalently \cref{eq:lll_rom_pz1,eq:lll_rom_mz1,eq:lll_rom_px1}) by expressing them in the matrix form as:
\begin{equation}
    \label{eq:matrixpde}
    \deriv{X}{t}=AX+b \quad \,,
\end{equation}
where we define the solution vector $X(t)=\left[\vdz(t),\vpz(t),\vpx(t)\right]$ (or as a function of $\vecm$ and $\vecp$ as $X(t)=\left[\vmz(t),\vpz(t), \vpx(t)\right]$) with the initial condition given by, 
\begin{flalign}
    \label{eq:initialconditionX}
    X_0=\begin{bmatrix}\vdz(t=0) \\ \vpz(t=0) \\ \vpx(t=0) \end{bmatrix}
    =
    \begin{bmatrix} 1 \\ -1 \\ 0 \end{bmatrix}
    &\quad
    \left( X_0=\begin{bmatrix}
    \vmz(t=0) \\
    \vpz(t=0) \\
    \vpx(t=0)
    \end{bmatrix}
    =\begin{bmatrix}
    -1 \\
    -1 \\
    0
    \end{bmatrix}
    \right) \, ,
\end{flalign}
and the homogeneous matrix $A$ and the forcing vector $b$ for the charge and optical channel (or the magnetization and optical) are given by:
\begin{flalign}
    \label{eq:matrixA_vectorb}
    A=\begin{bmatrix}
    -\ccol  & -\ccol & 0 \\
    0       &     0  & \Omega \\
    -\dfrac{\Omega}{2} & -\Omega & -\dfrac{\ccol}{2}-\pdeph
    \end{bmatrix}
    &\quad
    \left( 
    A=\begin{bmatrix}
    -\ccol  & \ccol & \dfrac{\Omega}{2} \\
    0       &     0  & \Omega \\
    \Omega & -\dfrac{3}{2}\Omega & -\dfrac{\ccol}{2}-\pdeph
    \end{bmatrix}
    \right) \, , \\
    \label{eq:vectorb}
    b&=\begin{bmatrix}
    0 \\ 0 \\ \dfrac{\Omega}{2}
    \end{bmatrix} \, .
\end{flalign}
The solution of the inhomogeneous system of equation can be expressed as:
\begin{equation}
    \label{eq:varprinciple}
    X(t)=\exp{\left(tA\right)}\left(X_0+\int_0^t{\exp{\left(-sA\right)}b(s)\mathrm{d}s}\right) \, .
\end{equation}
We first focus on the solution of the homogeneous problem, when no forcing factor is applied. It is possible to transform the general matrix $A$ using a linear map $\Tmap$ in $\mathbb{R}^3$ such that $B=\Tmap^{-1}A\Tmap$ such that $\exp{(A)}=\Tmap \exp{(B)}\Tmap^{-1}$ where $B$ is the canonical form of the matrix where the main diagonal is composed of the eigenvalues of $A$, and the solution vector is given in the Jordan canonical space is given by $y=\Tmap^{-1}X$. To put the system in the canonical form, we choose $\Tmap$ to be a matrix whose columns are the eigenvectors of $A$. If $A$ has distinct eigenvalues, the solution of the homogeneous problem can be expressed as,
\begin{equation}
    \label{eq:canonicical_diffeeigen}
    y(t)=\begin{bmatrix}
    y_1\exp{\left(\lambda_1 t\right)}  \\
    y_2\exp{\left(\lambda_2 t\right)} \\
    y_3\exp{\left(\lambda_3 t\right)} 
    \end{bmatrix} \, ,
\end{equation}
where $y_i$ with $i=1,2,3$ are the components of the canonical solution vector $y$. From \cref{eq:canonicical_diffeeigen} we can see that the the canonical solution can lead to two different regimes: (i) the system relaxes to 0 exponentially if the 3 eigenvalues are all real and negative (incoherent regime), (ii) the system relaxes to zero by a combination of oscillations and exponential decay if one of the eigenvalue is real and negative and the other two are complex conjugate (coherent regime). The particular type of behaviour is a function of the intensity of the optical excitation $\Omega$ and the dephasing $\pdeph$.

 For simplicity, we use the timescale in the units of $1/\ccol$, thus,
 \begin{flalign}
     \label{eq:tau}
     \tau&=t\ccol \, ,\\
     \label{eq:omega}
     \omega&=\dfrac{\Omega}{\ccol} \, ,\\
     \label{eq:gamma_asterisk}
     \gamma^*&=\dfrac{\pdeph}{\ccol} \, ,\\
     \label{eq:gamma_t}
     \gamma_t&=\dfrac{1}{2}+\gamma^* \, ,
 \end{flalign}
 and the eigenvalues of the rescaled Hamiltonian are used to determine the boundary between the incoherent and the coherent regime. For $\gamma_t=1/2$, the eigenvalue problem has a simple analytical solution,
 \begin{equation}
     \label{eq:eigenv_nopdeph}
     \lambda_1=\frac{1}{2}\qquad\lambda_{2,3}=-\frac{1}{2}\left(1\pm\sqrt{1-4\Omega^2}\right) \, .
 \end{equation}

 The system is driven into a critical regime at $\omega=1/2$ (i.e. three-fold degenerate state with real eigenvalue), and it is driven into a coherent regime for $\omega>1/2$.
   
  \begin{figure}
     \centering
     \includegraphics[width=\textwidth]{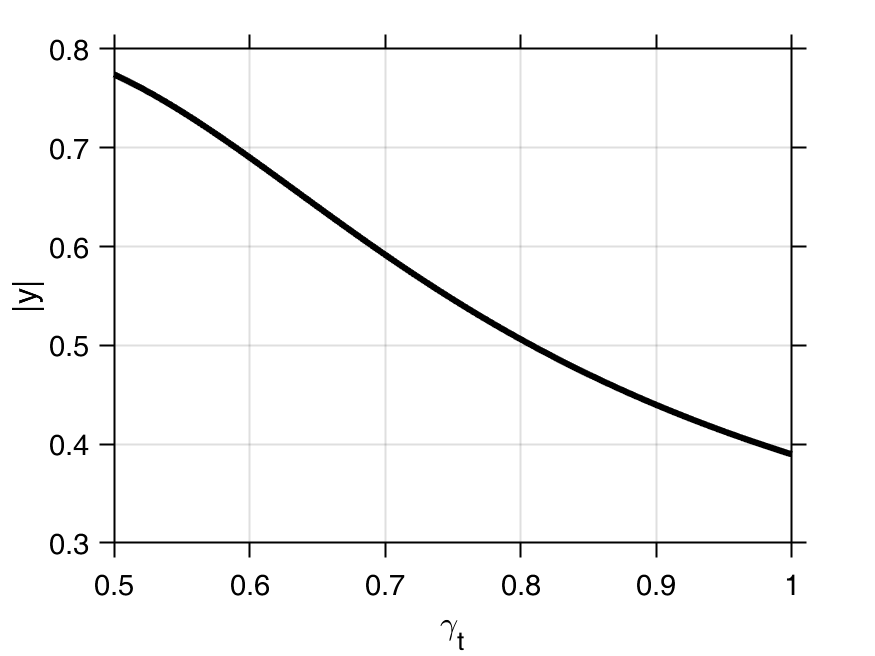}
     \caption{Module of the complex component $y_0$ in the canonical space associated with the complex eigenvector $\lambda$ as a function of $\gamma_t$ for $\omega=0.8$. }
     \label{fig:moduley}
 \end{figure}
 
\begin{figure}
     \centering
     \includegraphics[width=\textwidth]{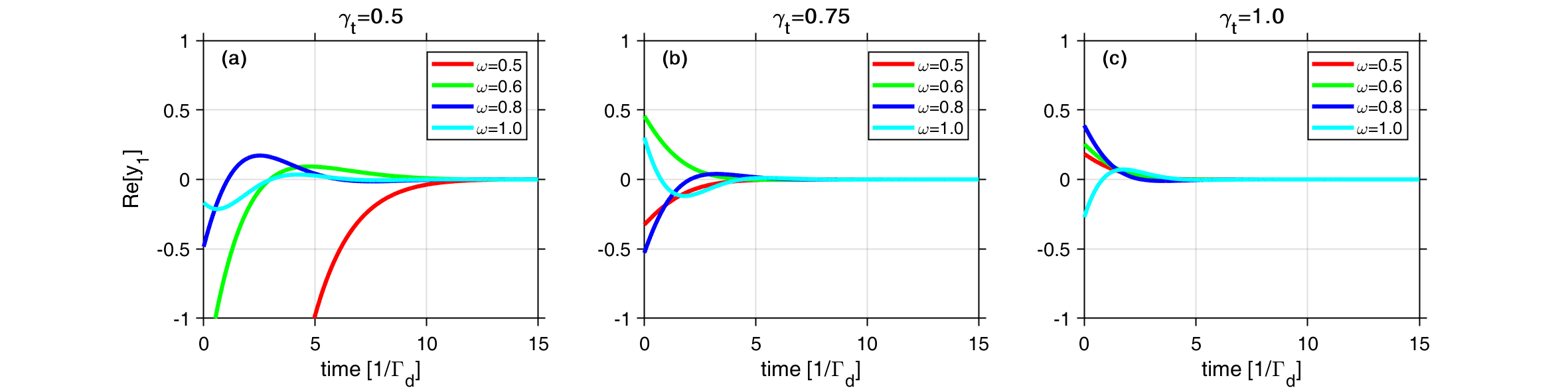}
     \caption{Dynamics of the real component of y in the canonical space for different values of the reduced optical excitation $\omega$. The results are shown for (a) no dephasing ($\gamma_t=0.5$), (b) $\gamma_t=0.75$, and (c) $\gamma_t=1.0$. }
     \label{fig:yspace}
 \end{figure}

 For $\gamma_t>1/2$, no similar critical regime is  observed. Increasing the pure dephasing $\gamma^*$, the value of the optical excitation  $\omega_{i,\mathbf{max}}$, for which the system goes from the incoherent regime to a coherent one, decreases almost linearly with $\gamma_t$. For $\gamma_t=1$ no incoherent regime is present. However, by studying the solution of the homogeneous problem in the $y$ space (\cref{fig:yspace}), we notice that for $\gamma_t>1/2$ (i.e. $\gamma^*>0$) even if the eigenvalues are complex (coherent excitation), the oscillation generated by the imaginary parts of the eigenvalue are rapidly quenched. Thus, a large value of $\gamma^*$ rapidly damps the coherence channel $\vpx$ that is at the origin of the coherent oscillation. This behaviour can be understood by observing the complex eigenvalue
 \begin{equation}
     \label{eq:complex_eignevalue}
     \lambda=-\alpha+ i\beta \, ,
 \end{equation}
 where $\alpha$ is the real and $\beta$ is the imaginary part of the eigenvalue. In the canonical space, the real part of the solution in the coherent regime can be expressed as
 \begin{equation}
     \label{eq:solutyspace}
     \mathrm{Re}[y]=\left(\mathrm{Re}[y_0]\cos{(\beta t)}-\mathrm{Im}[y_0]\sin{(\beta t)}\right)\exp(-\alpha t)
 \end{equation}
 
\begin{figure}
     \centering
     \includegraphics[width=\textwidth]{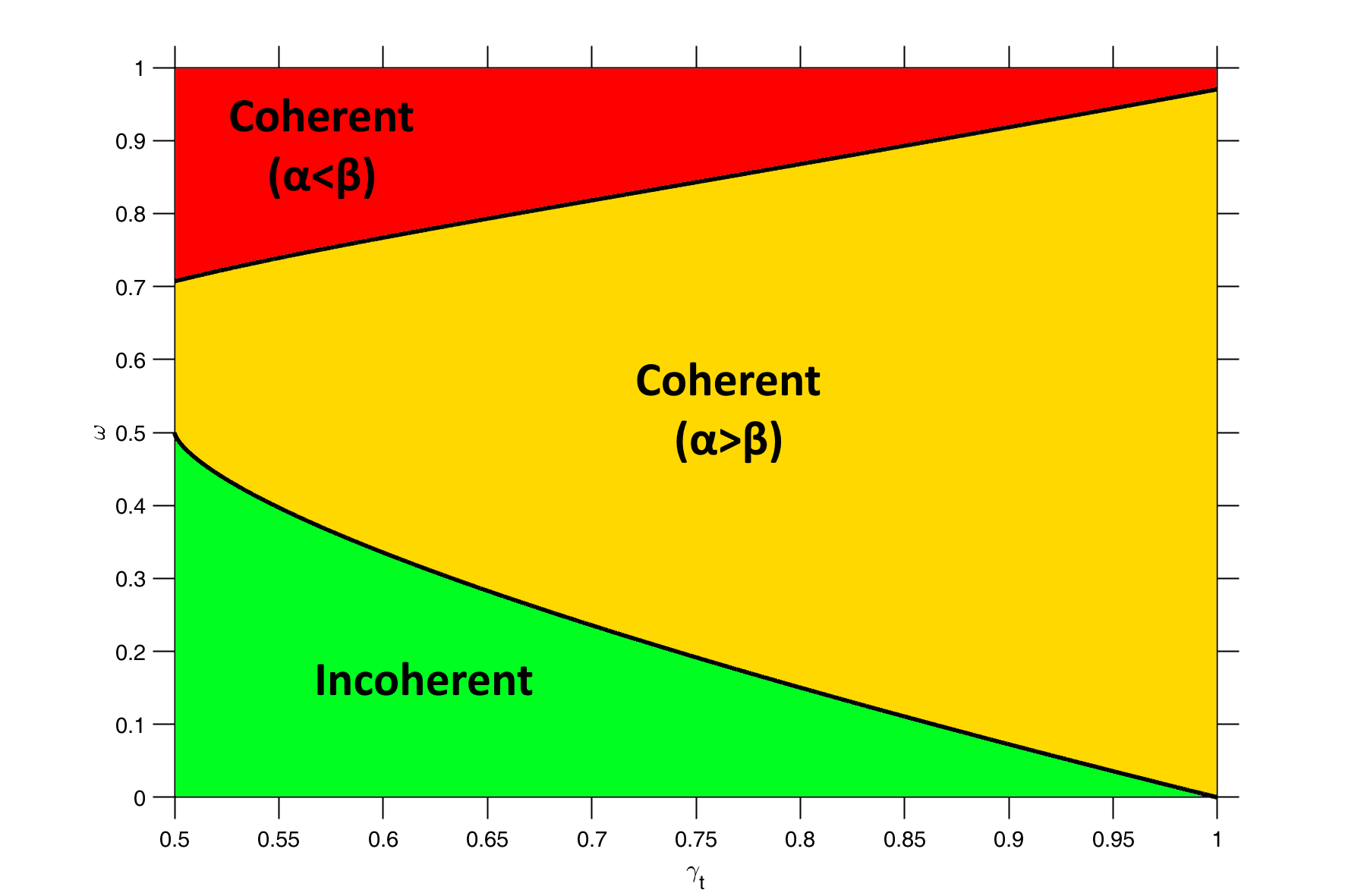}
     \caption{The incoherent regime (green area) is defined by real eigenvalue. For the coherent regime we define a weak oscillation region (yellow area) where the contribution of the decay is dominant, $\alpha>\beta$ and a strong oscillation region (red area) where the contribution of the oscillation is dominant, $\alpha<\beta$.  }
     \label{fig:regimestudy}
 \end{figure}

 When $\beta$ is small compared to $\alpha$, $y$ relaxes in a single oscillation. Moreover, the module of the complex component of $y_0$ associated with the complex eigenvalue $\lambda$, decreases with $\gamma_t$ as shown in \cref{fig:moduley}. This means that not only the number of oscillations necessary to reach the equilibrium, but also the amplitude of such oscillations is reduced when $\gamma_t$ increases. This allows us to define two regions in the coherent regime: (i) a weak oscillation regime ($\alpha>\beta$), and (ii) a strong oscillation regime ($\alpha<\beta$). When no dephasing is present ($\gamma_t=1/2$), the critical regime between the coherent and incoherent region can be found analytically by imposing in \cref{eq:eigenv_nopdeph} the condition $\mathrm{Re}\left[\lambda_2\right]=\mathrm{Im}\left[\lambda_2\right]$ that leads to $\omega=1/\sqrt{2}$. For $\gamma_t>1/2$, the transition between weak and strong oscillation regimes is obtained numerically. The three regimes of interests are shown in \cref{fig:regimestudy}.
 
 To study the dynamics of the inhomogeneous problem, we need to obtain the contribution of the forcing vector in the homogeneous solution. Since the forcing vector $b^{(1)}=T^{-1}b$ is constant, its contribution can be easily obtained for distinct eignevalues as,
 \begin{equation}
     \label{eq:inhom_contribution}
     \exp{\left(tB\right)}\int_{0}^{t}{\exp{\left(-sB\right)}b^{(1)}}=\begin{bmatrix}
     \dfrac{e^{\lambda_1 t}-1}{\lambda_1}b^{(1)}_{1} \\
     \dfrac{e^{\lambda_2 t}-1}{\lambda_2}b^{(1)}_{2} \\
     \dfrac{e^{\lambda_3 t}-1}{\lambda_3}b^{(1)}_{3} 
     \end{bmatrix} \, ,
 \end{equation}
 where $b^{(1)}_i$ with $i=1,2,3$ are the components of the forcing vector. By adding \cref{eq:inhom_contribution} to \cref{eq:canonicical_diffeeigen} and applying the mapping operator $\Tmap$ to the solution, we obtain the solution of the resonant optical model for the charge and the optical polarization $X(t)=\Tmap y(t)$. If we set the time to $t\rightarrow\infty$ to let the transitory disappear, the homogeneous contribution given in  \cref{eq:canonicical_diffeeigen} goes to zero and in the steady state we obtain
 \begin{equation}
     \label{eq:steadystate}
     \begin{array}{lcr}
     y_{\mathrm{steady}}=\begin{bmatrix}
     \dfrac{-b^{(1)}_{1}}{\lambda_1} \\
     \dfrac{-b^{(1)}_{2}}{\lambda_2}  \\
     \dfrac{-b^{(1)}_{3}}{\lambda_3}  
     \end{bmatrix}
     & \rightarrow  &
     X_{\mathrm{steady}}=T\cdot y_{\mathrm{steady}}=\begin{bmatrix}
     -1 \\
     1 \\
     0 
     \end{bmatrix}
     \end{array}
     \, .
 \end{equation}
 Equation (\ref{eq:steadystate}) shows that the optical excitation is able to empty $\stb$ ($\rho_{22,\mathrm{steady}}=0$) initially occupied and to fill initially empty $\sta$ ($\rho_{22,\mathrm{steady}}=1$). Thus, the magnetization population (see \cref{eq:magnetizationpopulation}) is reversed in the process.
 
 To estimate the efficiency of the reversal, we define the reversal magnetization rate $\revr$  as the inverse of the reversal time $\revt$. The reversal time is chosen as the time for which the value of the magnetization becomes $\vmz(\revt)=0.9$. We want to study how the reversal rate and dynamics are influenced by the the optical excitation $\omega$ and the pure dephasing $\gamma^*=\pdeph/\ccol$.
 \begin{figure}
     \centering
     \includegraphics[width=\textwidth]{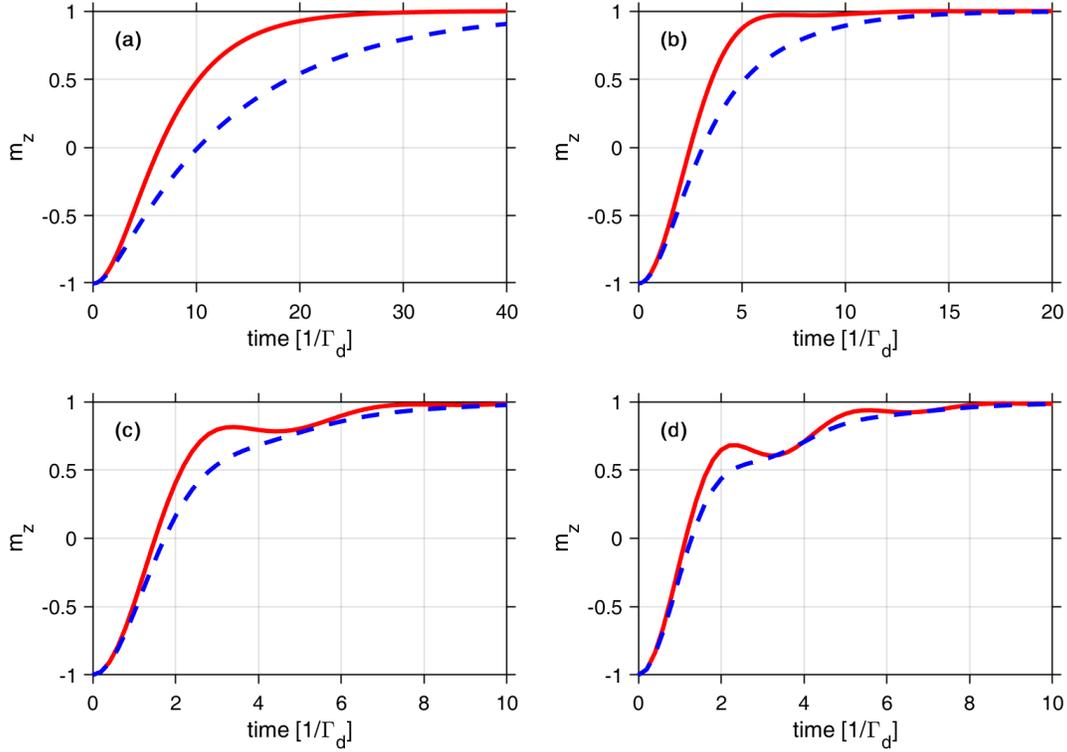}
     \caption{Magnetization dynamics as a function of time for: (a) $\omega=0.4$, (b) $\omega=0.9$, (c) $\omega=1.5$, and $\omega=2.0$. The solid red line are computed for $\gamma^*=0.0$ and the blue dashed ones for $\gamma^*=0.5$.}
     \label{fig:dynamics}
 \end{figure}
 
\begin{figure}
     \centering
     \includegraphics[width=\textwidth]{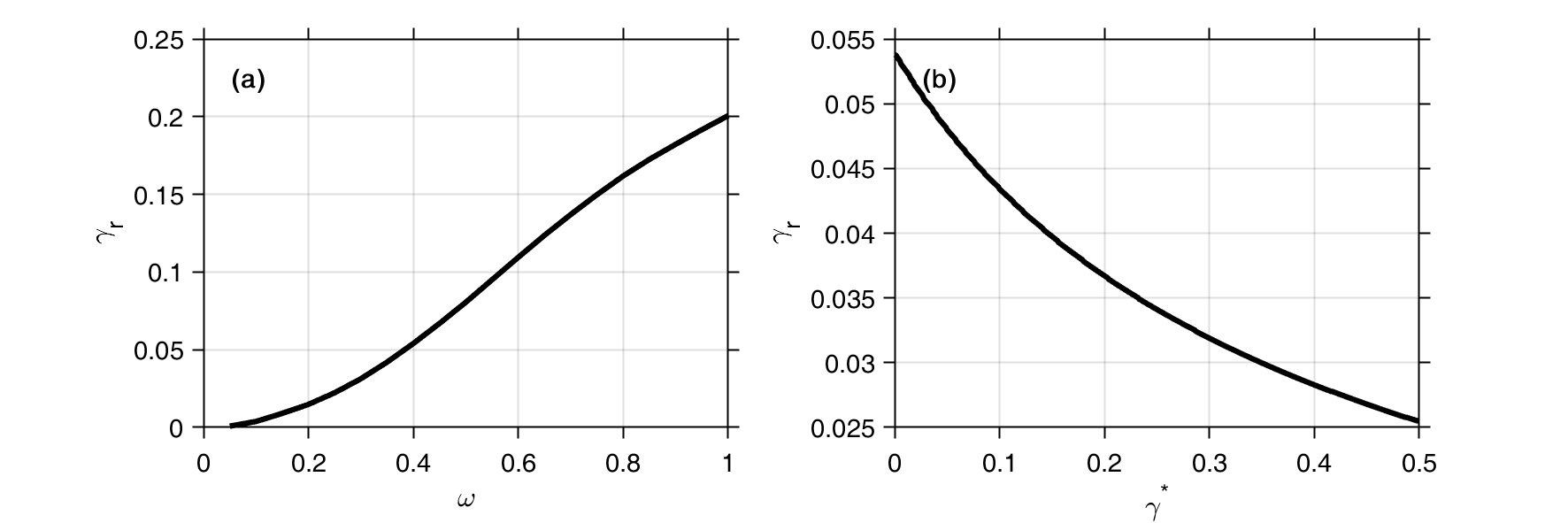}
     \caption{Reversal rate as a function of (a) $\omega$ for $\gamma^*=0$, and (b) as a function of $\gamma^*$ for $\omega=0.4$}
     \label{fig:dependency_omega_gamma}
 \end{figure}

For low values of $\omega<1$, no significant oscillations of the magnetization are observed in either the incoherent regime, $\omega<1/2$, (\cref{fig:dynamics}a), or the coherent one (\cref{fig:dynamics}b). Oscillatory behaviour becomes dominant for $\omega\gg 1$ (\cref{fig:dynamics}c-d). Increasing the optical excitation increases the reversal rate up to $\omega=1$ (\cref{fig:dependency_omega_gamma}a). When the optical excitation exceeds the Coulomb collapse (i.e. $\omega>1$), the reversal rate plateaus, and the efficiency of the reversal is limited by Coulomb collapse (i.e. population inversion and lasing). The presence of dephasing $\gamma^*$ acting on the coherence of the polarization channel leads to slower reversal rates for $\omega\le 1$ (\cref{fig:dependency_omega_gamma}), while it only quenches the oscillatory behaviour during the plateau $\gamma^*$ (\cref{fig:dynamics}c-d).

\section{\label{sec:incoherent} Incoherent Excitation Approximation}

In many practical cases, we are interested in studying phenomena of optical reversal when  a system is excited in the incoherent regime ($\lambda_i\in\mathbb{R} \ \forall i=1,2,3$) or a coherent regime with weak oscillatory behaviour ($\mathrm{Re}{[\lambda]}\gg\mathrm{Im}{[\lambda]}$). In this section we introduce an approximation to the LLL model based on the assumption of an incoherent excitation, and we compare the results with the model we developed in the previous sections. Under the incoherent approximation, we can reduce the number of unknowns from 8N to 4N, only including the population $\vpz$ and the magnetization vector $\vecm$. This reduces the amount of memory and operations required to solve the system numerically.
Moreover, in the incoherent approximation, the optical excitation is represented by an effective pumping $G$ between $\stb$ and $\stc$, which can be calculated from first principles \cite{scheid2019ab}. The value of the optical pumping $G$ can be obtained by using the Fermi golden rule:
\begin{equation}
\label{eq:fermigoldenrule}
G=\frac{2\pi}{\hbar}\left|\braket{2|H^\prime|3}\right|^2 \rho(E_{23})\, ,
\end{equation}
where $H^\prime$ is the perturbation of the Hamiltonian, and $\rho(E_{23})$ is the joint density of states per unit of energy between $\stb$ and $\stc$. 

\begin{figure}
    \centering
    \includegraphics[width=0.8\textwidth]{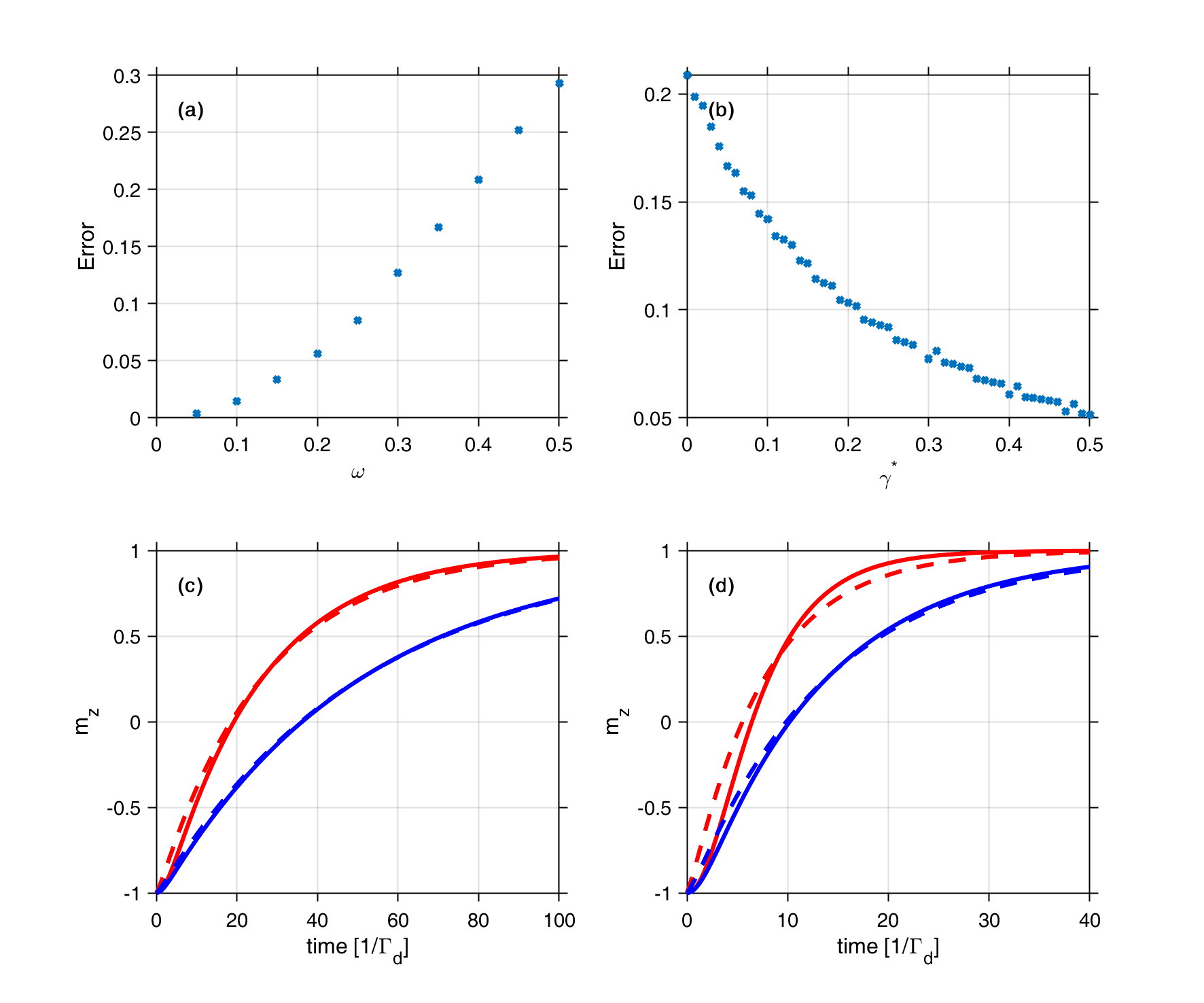}
    \caption{ Error introduced by the incoherent approximation (a) as a function of $\omega$ for $\gamma^*=0$ and (b)  as a function of $\gamma^*$ for $\omega=0.4$. The magnetization dynamics is shown for the the LLL model (solid lines) and the incoherent approximation (dashed lines) for (c) $\omega=0.2$ and (d) $\omega=0.4$. The red lines show the result for $\gamma^*=0$ and the blue lines for $\gamma^*=0.5$.}
    \label{fig:comparison}
\end{figure}

We start with the resonant optical model defined in \cref{eq:lll_rom_dz,eq:lll_rom_pz,eq:lll_rom_px}. Since the dependence of  the coherence $\vpx$ only appears explicitly in the optical polarization $\vpz$ (\cref{eq:lll_rom_pz}), we can rewrite the optical population $\vpz$ for the resonant optical model equations in terms of the population using the operator method \cite{tzou2014},
\begin{equation}
    \label{eq:lll_rom_pz2}
    \secderiv{\vpz}{\tau}+2\zeta\omega\deriv{\vpz}{t}+\omega^2\vpz=\dfrac{\omega^2}{2}\left(1-\vdz\right) \, ,
\end{equation}
where $\zeta=(1/2+\gamma^*)/(2\omega)$ is the damping ratio. When $\zeta\gg1$ (i.e. $\omega\ll 1+2\gamma^*$), $\vpz$ behaves as an over-damped oscillator and no coherence magnon excitation is observed during the magnetization reversal. In this case, we can neglect the contribution of the slowly varying second derivative and rewrite an equation for the $\vpz$ as
\begin{equation}
    \label{eq:lll1_rom_pz}
    \deriv{\vpz}{\tau}=-\dfrac{\omega^2}{1+2\gamma^*}\left(2\vpz+\vdz-1\right) \, .
\end{equation}
We will refer to the model described by eqs. \cref{eq:lll_rom_dz,eq:lll1_rom_pz} LLL-1. The term $\omega^2/(1+2\gamma^*)$ can be understood as an effective pumping coefficient exciting the population $\stb$ into $\stc$. The excitation is followed by a fast decay without change of spin into $\sta$ due to the Coulomb collapse (\cref{eq:lll_rom_dz}). 
We define the pumping parameter G in the form:
\begin{equation}
    \label{eq:pumping_coefficient}
    G=\frac{\Omega^2}{\ccol+2\pdeph}\, .
\end{equation}
By using \cref{eq:magnetizationpopulation}, we can rewrite the LLL-1 form of the resonant optical model in terms of the population $\vpz$ and the magnetization vector $\vmz$ in terms of the time $t$ instead of the reduced time $\tau=t\ccol$ as:
\begin{flalign}
    \label{eq:lll1_rom_mz}
    \deriv{\vmz}{t}&=\left(\ccol-G\right)\left(\vpz-\vmz\right)-\frac{G}{2}\left(\vpz-1\right)\approx
    \ccol\left(\vpz-\vmz\right)-\frac{G}{2}\left(\vpz-1\right)\, , \\
    \label{eq:lll1_rom_pz3}
    \deriv{\vpz}{t}&=-2G\left(\vpz-\vmz\right)-G\left(\vpz-1\right)\approx-G\left(\vpz-1\right)\, ,
\end{flalign}
where the last approximation is obtained under the assumption $G\ll\ccol$, valid in the incoherent regime and for large values of $\pdeph$. The approximation in \cref{eq:lll1_rom_mz,eq:lll1_rom_pz3} is equivalent to the result obtained by introducing the optical excitation in master equation as an incoherent pumping through the dissipative Lindblad operator $\lindblad{G}$:
\begin{equation}
    \label{eq:pumpingoperator}
    \lindblad{G}=
    \begin{bmatrix}
    0 & 0 & 0 \\
    0 & 0 & 0 \\
    0 & \sqrt{G} & 0
    \end{bmatrix}\, .
\end{equation}
To study the error introduced by using the incoherent approximation in \cref{eq:lll1_rom_mz,eq:lll1_rom_pz3}, we introduce the parameter
\begin{equation}
    \label{eq:error}
    \mathrm{Error}=\dfrac{\left|\gamma_r-\gamma_r^\prime\right|}{\gamma_r} \, ,
\end{equation}
where $\gamma_r^\prime$ is the reversal rate obtained using the incoherent approximation and $\gamma_r$ is the reversal time obtained by using \cref{eq:lll_rom_mz1,eq:lll_rom_pz}. The two models show a similar behavior in the incoherent regime (\cref{fig:comparison}a) and for large dephasing (\cref{fig:comparison}b). The results show that in the incoherent regime, which we define as $\omega\ll 1/2$, the results obtained using the LLL-1 and the LLL model lead to similar dynamics (\cref{fig:comparison}c). For larger values of $\omega$ , close to the transition between the coherent and incoherent regimes, small differences in the dynamics are observed  (\cref{fig:comparison}d). Increasing $\pdeph$, increases the range of validity of the model.

In ferromagnetic materials, such as FePt and CoPt, the decay rate is estimate to be of the order of $\ccol\approx 1 \mathrm{fs^{-1}}$, which is based on the lifetime of the photo-excited electrons in Fe \cite{Schmidt2010} and the $\pdeph$ is proportional to the electron temperature reached by the laser. The peak optical power used in the HD-AOS experiments is of the order of $P=0.01-0.04 \mathrm{TW/cm^{2}}$. If we assume a dipolar transition moment on the order on $\mue\approx 1 \mathrm{e\text{\AA}}$, we obtain a Rabi frequency for $\lhc$ on the order of $\Omega=0.02-0.04$. In this range, the error introduced by using the LLL-1 model is negligible.

\section{\label{sec:conclusions} Summary}

We introduced a Landau-Lifshitz-Lambda (LLL) model, which is a theoretical framework to describe the excitation of magnetic materials using circular polarized light. The switching between magnetic states $\stb$ and  $\sta$ is meditated by an optically excited state $\stc$. The optical excitation drives the population of the system, initially in the $\stb$ state, to oscillate between the ground state and the excited state $\stc$. The presence of Coulomb collapse between $\stc$ and $\sta$ generates a leakage after every oscillation that reduces the population in $\stb$, increasing the population in  $\sta$, effectively reversing the magnetization.

The evolution in time of the optically excited states is described by a Lindblad master equation, where the optical excitation is introduced via the Hamiltonian as a Rabi frequency between the states $\stb$ and $\stc$. The dissipation is introduced in the system by using two Lindblad operators: one to describe the contribution of the electron-electron Coulomb collapse, and one to describe dephasing. The contribution of the dissipation in the magnetic sector are neglected since the timescale of the optical relaxation $\tau_{opt}$ is order of magnitude smaller than the timescale of the magnetic relaxation $\tau_{m}$ (i.e. $\tau_{opt}\ll\tau_{m}$). The elements of the density matrix, representing the 3 states, are transformed into Bloch vectors to obtain a model for the magnetization dynamics for ferromagnetic materials subjected to optical and magnetic excitation ready to be used with existing micromagnetic modeling software. The LLL model combines the precessional motion of the magnetization described by the Landau-Lifshitz theory, with the response of a three level $\Lambda$ system.

We studied the effect of the optical excitation at resonance and in the absence of an applied field as a testbed. We demonstrated that the coherent Rabi excitation coupled with a fast electronic decay is able to produce a complete optical reversal of the magnetization in several hundred femtoseconds. By studying the eigenvalues of the resonant optical model, we have been able to find 3 difference regimes in the system: A purely incoherent one, where all the eigenvalues are real and the magnetization is able to switch from down to up exponentially without oscillations; a weak oscillatory regime, where the excitation is effectively coherent (complex eigenvalues), but the magnitude of the oscillation behaviour is negligible; and a strong oscillatory regime.

Under the assumption of incoherent or weak oscillatory regime (i.e. an effective decay rate much greater than the Rabi frequency), the model can be expressed in the form of an effective pumping rate $G$. This approximation, usually valid for ferromagnetic materials, such as FePt or CoPt, leads to a simpler LLL-1 model. This model is accurate in describing the magnetization dynamics in the range of laser power and applied fields commonly used in the AOS experiments, easy to be integrated with existing micromagnetic codes, and may allow obtaining the pumping rate $G$ based on first principles calculations.

In the LLL model, the effect of the optical excitation emerges naturally from the equation of motion of the magnetization. This makes the models suitable to describe the magnetization dynamics and switching in a wide range of magnetic materials and properties subject to circularly polarized optical excitation. The model could also be used to describe the helicity dependent domain wall motion (HD-DWD) recently observed in Co/Pt thin films when excited by optical pulses with low fluency \cite{Medapalli2017,Quessab2018}. It can also be extended to describe effects of any kind of elliptically polarized light.

\section{\label{sec:aknowledgment} Acknowledgments}

This work was funded by National Science Foundation. The authors thank Professor Eric E. Fullerton, Professor Stéphane Mangin, Dr. Rajasekhar Medapalli, and Philippe Sched for the discussions on experimental and simulation data. We wish to warmly thank Professor Lu Jeu Sham for fruitful discussions and for his contributions in developing the framework of the optical excitation model.

\appendix

\section{\label{sec:app_vector} Vector representation of the density of operator for the $\Lambda$ system}

The equation of motion of the density matrix \cref{eq:lindblad} is based on the formulation of the Lindblad operators \cite{Lindblad1976} as the generators of the semigroup that is not time reversible. The evolution operators of the three state Hamiltonian form a unitary group SU(3) that is time reversible. The dissipation terms would drive the system only in the forward time direction and hence, the formulation is to utilize the semigroup of transformations which reflects the time direction for dissipative effects. The generators of the transformations are called the Lindblad operators. Explicit formulation of similar operators in N-state systems, especially N=2, are given in Ref. \cite{Gorini1976} and are clearly explained in Ref. \cite{Schumacher2010}. These generators simulate the dissipative and dephase effects which may be derived from the quantum dynamics of the quantum system plus its environment defined as a quantum system. When the environment system is traced out of the evolution operators, the rate results are given in terms the Kraus operators \cite{Kraus}, which are modeled by the Lindblad operators. We model the equation of motion of the system, \cref{eq:lindblad}, as the dynamics of the open system by the formalism of Lindblad \cite{Lindblad1976}.

The population decays (i.e., longitudinal relaxations) in the $\Lambda$ system shown in \cref{fig:scheletonprocess} are treated as two separate processes since the decay in the optical sector is due to spin-flip processes and the decay in the charge sector is assumed to be from the inelastic electron-
electron scatterings without spin flips. This idea is based on the experimental measurements of the fast decay. MOKE contrast shows a remanent magnetization leading to the inference of instantaneous “Stoner gap collapse” \cite{Cheskis2005}. The dominant electron-electron scattering rate is proportional to the number of unoccupied electronic states into which excited electrons can scatter \cite{Knoesel1996,Aeschlimann1997}. Hence, the relaxation from state 3 to state 1 is modeled by the Lindblad operator for dissipation, known as the longitudinal relaxation rate,
\begin{equation}
    \label{eq:operatorcoulombcollapse}
    \lindblad{d}=\begin{bmatrix}
    0 & 0 & \sqrt{\ccol} \\
    0 & 0 & 0 \\
    0 & 0 & 0 
    \end{bmatrix} \, ,
\end{equation}
where  $\sqrt{\ccol}$ leads to the fast relaxation in the decay sector $\stc\rightarrow\sta$. The electronic decay rate is shown by experiments on fast optical demagnetization to be in the femtosecond range \cite{Cheskis2005,Aeschlimann1997,Rhie2003}. This Lindbland operation simultaneously produces also a decoherence effect on the off-diagonal terms of the density matrix with half the decaying rate ($\ccol/2$).

In the optical sector, the spontaneous radiative emission from $\stc\rightarrow\stb$ is negligible compared with the usual laser power and the spin-flip decay is assumed to be weaker than the optical excitation, and its contribution is neglected in the model. While the  population decay may be considered as negligible, the pure dephasing without population decay between these two states may be driven by the distribution of low-lying states around the two macrospin states. In the magnetization dynamics, the phase between the two magnetization states is randomized to a certain extent by the distributions of the two states, caused by the spin waves or Stoner-type low energy collective spin excitation \cite{Galanakis2012, Glazer1984}. The Lindblad operator for the pure decoherence (i.e., dephasing) applies to both sectors,
\begin{equation}
    \label{eq:operatorpuredephasing}
    \lindblad{2}=\begin{bmatrix}
    \sqrt{\pdeph_{1}} & 0 & 0 \\
    0 &  \sqrt{\pdeph_{2}} & 0 \\
    0 & 0 &  -\sqrt{\pdeph_{1}}- \sqrt{\pdeph_{2}} 
    \end{bmatrix} \, .
\end{equation}
For simplicity, we assume that the two magnetization states are completely equivalent and so, $\pdeph_{1}=\pdeph_{2}$. The diagonal and traceless form of this operator will produce no population change in the states but only their relative phases,
\begin{equation}
    \label{eq:effectdephasing}
    \left(\lindblad{2}\dens\lindblad{2}^\dagger-\dfrac{1}{2}\acommut{\lindblad{2}^\dagger\lindblad{2}}{\dens}\right)
    \rightarrow
    -\begin{bmatrix}
    0   & 0 & -\dfrac{9}{2}\pdeph_{1}\rho_{13}\pdeph \\
    0   & 0 & -\dfrac{9}{2}\pdeph_{1}\rho_{23} \\
    -\dfrac{9}{2}\pdeph_{1}\rho_{31}   & -\dfrac{9}{2}\pdeph_{1}\rho_{32} & 0 \\
    \end{bmatrix} \, .
\end{equation}
The pure decoherence rate is given by $\pdeph=-\frac{9}{2}\pdeph_{1}$. We have neglected the direct decay and decoherence between the two ground macrospin states, because the independent magnetic
relaxation times are much longer than the time scale of the optoelectronic processes.

\section{\label{sec:app_motion} Equation of motion}

We consider the mathematical structure that leads to the decomposition of the density matrix into the components in \cref{sec:motion}. Its difference from the standard decomposition as generalized Bloch vectors \cite{Bertlmann2008,Kimura2003} and the physical consequences is detailed. The density matrix in \cref{eq:densitymatrix} can be rewritten in terms of the linear combination of nine Hermitian operators,
\begin{equation}
    \label{eq:densityhermitian}
    \dens=\dfrac{1}{2}\left[\vmx\sutr{12}{x}+\vmy\sutr{12}{y}+
    \vpx\sutr{23}{x}+\vpy\sutr{23}{y}+
    \vdx\sutr{13}{x}+\vdy\sutr{13}{y}+
    \vdz\sutr{13}{z}+\vpz\sutr{23}{z}+
    \hat{\mathrm{I}}_{12}{z}
    \right] \, .
\end{equation}
The matrix representations of the Hermitian operators associated with the polarization components $\vmx,\vmy,\vpx,\vpy,\vdx,\vdy$  are, respectively,
\begin{equation}
    \label{eq:suoperators}
    \begin{array}{ccc}
    \sutr{12}{x}=
    \begin{bmatrix}
    0 & 1 & 0 \\
    1 & 0 & 0 \\
    0 & 0 & 0 \\
    \end{bmatrix} \, ,
    &
    \sutr{12}{y}=
    \begin{bmatrix}
    0 & -i & 0 \\
    i & 0 & 0 \\
    0 & 0 & 0 \\
    \end{bmatrix} \, ,
    &
    \sutr{23}{x}=
    \begin{bmatrix}
    0 & 0 & 0 \\
    0 & 0 & 1 \\
    0 & 1 & 0 \\
    \end{bmatrix} \, ,
         \\
    \sutr{23}{y}=
    \begin{bmatrix}
    0 & 0 & 0 \\
    0 & 0 & i \\
    0 & -i & 0 \\
    \end{bmatrix} \, ,
          &
    \sutr{13}{x}=
    \begin{bmatrix}
    0 & 0 & 1 \\
    0 & 0 & 0 \\
    1 & 0 & 0 \\
    \end{bmatrix} \, ,
          &
    \sutr{13}{y}=
    \begin{bmatrix}
    0 & 0 & i \\
    0 & 0 & 0 \\
    -i & 0 & 0 \\
    \end{bmatrix} \, ,
    \end{array}
\end{equation}
which are the same of the generators in the group SU(3) \cite{GellMann1962}. Our diagonal matrices associated with the coefficients $\vdz,\vpz,1$ are
\begin{equation}
    \label{eq:populationnoperators}
    \begin{array}{ccc}
    \sutr{13}{z}=
    \begin{bmatrix}
    -1 & 0 & 0 \\
    0 & 0 & 0 \\
    0 & 0 & 1 \\
    \end{bmatrix} \, ,
    &
    \sutr{23}{z}=
    \begin{bmatrix}
    0 & 0 & 0 \\
    0 & -1 & 0 \\
    0 & 0 & 1 \\
    \end{bmatrix} \, ,
    &
    \hat{\mathrm{I}}_{12}^{z}=
    \begin{bmatrix}
    1 & 0 & 0 \\
    0 & 1 & 0 \\
    0 & 0 & 0 \\
    \end{bmatrix} \, .
    \end{array}
\end{equation}
These nine operators are the Pauli operators and a unit operator of the spin $1/2$ subspaces and may be regarded as a linearly independent set of basis states in a vector group. The diagonal elements are not orthogonal but may be rendered so, for example, as the standard SU(3) generators \cite{GellMann1964}.


Our chosen generalized Bloch operators are under the restrictions of a Hermitian matrix representing a density matrix given by:
\begin{enumerate}
    \item The unit trace condition $\Tr{(\dens)}=1$, which is satisfied by the form of the diagonal elements in \cref{eq:densitymatrix} involving only the longitudinal components $\vdz,\vpz$ .
    \item The condition of a positive matrix leads to
    \begin{equation}
        \label{eq:condtrace}
        \vdz\le 1 \, , \qquad \vpz\le 1 \, , \qquad \vdz+\vpz\ge 0 \, , 
    \end{equation}
\end{enumerate}
The inequality of  $\Tr{(\dens^2)}\le1$ derived from the unit trace condition yields,
    \begin{equation}
        \label{eq:condtrace2}
        P^2+D^2+\vmx^2+\vmy^2+\vdz\vpz-\vdz^2-\vpz^2=1 \, , 
    \end{equation}
where $D^2,P^2$ are the squared magnitudes of the polarization vectors $\vecd,\vecp$.

\begin{figure}
    \centering
    \includegraphics{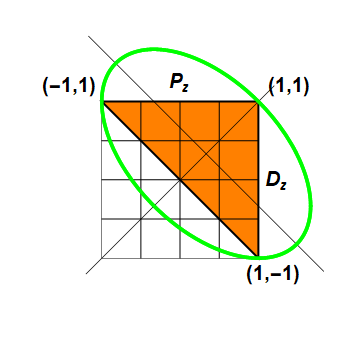}
    \caption{The polarization cross-section $(\vdz,\vpz)$  in the eight dimension polarization space. The ellipse centered at  $(1/3,1/3)$  with the major axis in the $(-1,1)$  direction and the minor axis along $(1,1)$  containsinside the region of  $\Tr{(\dens^2)}$. The valid density matrices are in the shaded region bound by the triangle with the vertices at the three pure states important to the optical processes, viz., the initial magnetization state 2 at $(-1,1)$ , the optically excited state 3 at $(1,1)$    and the final magnetization reversal state 1 at $(1,-1)$ .}
    \label{fig:triangulardomain}
\end{figure}

In the two-state system, the unit trace and positive requirements of the density matrix yields the same restriction as  $\Tr{(\dens^2)}\le1$ for the Block vectors within a unit sphere with the pure states on the surface. Such a coincidence does not occur in systems with more than two states since $\Tr{(\dens^2)}\le1$ is a weaker condition than $\Tr{(\dens)}=1$. This is seen in \cref{fig:triangulardomain}, where the triangular domain inside the three lines given by \cref{eq:condtrace} in the $\vdz,\vpz$ plane when all the transverse (i.e., x, y) components of the polarization vectors are zero. The states on the triangle border are pure states and those inside are mixed states. The ellipse domain in the same $\vdz,\vpz$ plane given by \cref{eq:condtrace2} contains but does not coincide with the triangular domain of legitimate states. This is similar to the sectional diagrams in the standard generalized Bloch vectors on p. 86 of Ref.~\cite{Mahler1998}, the difference being only the skewed basis of our density matrix. The relevant feature of the triangle in our case is that the three vertices $\left(\vdz,\vpz\right)=\left(-1,1\right),\left(1,1\right),\left(1,-1\right)$ correspond respectively to the initial magnetization state 2, the optically excited state 3, and the reversed magnetization state 1. The counterclockwise motion along the rims of the triangle is a possible dynamical path. 


\newpage
\medskip

\bibliography{bibliography.bib}

\end{document}